\documentclass[conference]{IEEEtran}
\IEEEoverridecommandlockouts
\usepackage{amsmath,amssymb,amsfonts}
\usepackage{algorithmic}
\usepackage{graphicx}
\usepackage{textcomp}
\usepackage[dvipsnames]{xcolor}
\usepackage{acronym}
\usepackage[sorting=none]{biblatex} 
\def\BibTeX{{\rm B\kern-.05em{\sc i\kern-.025em b}\kern-.08em
    T\kern-.1667em\lower.7ex\hbox{E}\kern-.125emX}}
\begin{document}

\acrodef{AI}[AI]{Artificial Intelligence}
\acrodef{ADC}[ADC]{Analog to Digital Converter}
\acrodef{ADEXP}[AdExp-I\&F]{Adaptive-Exponential Integrate and Fire}
\acrodef{AER}[AER]{Address-Event Representation}
\acrodef{AEX}[AEX]{AER EXtension board}
\acrodef{AE}[AE]{Address-Event}
\acrodef{AFM}[AFM]{Atomic Force Microscope}
\acrodef{AGC}[AGC]{Automatic Gain Control}
\acrodef{AMDA}[AMDA]{AER Motherboard with D/A converters}
\acrodef{ANN}[ANN]{Artificial Neural Network}
\acrodef{API}[API]{Application Programming Interface}
\acrodef{ARM}[ARM]{Advanced RISC Machine}
\acrodef{ASIC}[ASIC]{Application Specific Integrated Circuit}
\acrodef{AdExp}[AdExp-IF]{Adaptive Exponential Integrate-and-Fire}
\acrodef{BCM}[BMC]{Bienenstock-Cooper-Munro}
\acrodef{BD}[BD]{Bundled Data}
\acrodef{BEOL}[BEOL]{Back-end of Line}
\acrodef{BG}[BG]{Bias Generator}
\acrodef{BMI}[BMI]{Brain-Machince Interface}
\acrodef{BPTT}[BPTT]{Backpropagation Through Time}
\acrodef{BTB}[BTB]{band-to-band tunnelling}
\acrodef{CAD}[CAD]{Computer Aided Design}
\acrodef{CAM}[CAM]{Content Addressable Memory}
\acrodef{CAVIAR}[CAVIAR]{Convolution AER Vision Architecture for Real-Time}
\acrodef{CA}[CA]{Cortical Automaton}
\acrodef{CCN}[CCN]{Cooperative and Competitive Network}
\acrodef{CDR}[CDR]{Clock-Data Recovery}
\acrodef{CFC}[CFC]{Current to Frequency Converter}
\acrodef{CHP}[CHP]{Communicating Hardware Processes}
\acrodef{CMIM}[CMIM]{Metal-insulator-metal Capacitor}
\acrodef{CML}[CML]{Current Mode Logic}
\acrodef{CMOL}[CMOL]{Hybrid CMOS nanoelectronic circuits}
\acrodef{CMOS}[CMOS]{Complementary Metal-Oxide-Semiconductor}
\acrodef{CNN}[CCN]{Convolutional Neural Network}
\acrodef{COTS}[COTS]{Commercial Off-The-Shelf}
\acrodef{CPG}[CPG]{Central Pattern Generator}
\acrodef{CPLD}[CPLD]{Complex Programmable Logic Device}
\acrodef{CPU}[CPU]{Central Processing Unit}
\acrodef{CSM}[CSM]{Cortical State Machine}
\acrodef{CSP}[CSP]{Constraint Satisfaction Problem}
\acrodef{CV}[CV]{Coefficient of Variation}
\acrodef{DAC}[DAC]{Digital to Analog Converter}
\acrodef{DAS}[DAS]{Dynamic Auditory Sensor}
\acrodef{DAVIS}[DAVIS]{Dynamic and Active Pixel Vision Sensor}
\acrodef{DBN}[DBN]{Deep Belief Network}
\acrodef{DFA}[DFA]{Deterministic Finite Automaton}
\acrodef{DIBL}[DIBL]{drain-induced-barrier-lowering}
\acrodef{DI}[DI]{delay insensitive}
\acrodef{DMA}[DMA]{Direct Memory Access}
\acrodef{DNF}[DNF]{Dynamic Neural Field}
\acrodef{DNN}[DNN]{Deep Neural Network}
\acrodef{DOF}[DOF]{Degrees of Freedom}
\acrodef{DPE}[DPE]{Dynamic Parameter Estimation}
\acrodef{DPI}[DPI]{Differential Pair Integrator}
\acrodef{DRRZ}[DR-RZ]{Dual-Rail Return-to-Zero}
\acrodef{DRAM}[DRAM]{Dynamic Random Access Memory}
\acrodef{DR}[DR]{Dual Rail}
\acrodef{DSP}[DSP]{Digital Signal Processor}
\acrodef{DVS}[DVS]{Dynamic Vision Sensor}
\acrodef{DYNAP}[DYNAP]{Dynamic Neuromorphic Asynchronous Processor}
\acrodef{EBL}[EBL]{Electron Beam Lithography}
\acrodef{EDVAC}[EDVAC]{Electronic Discrete Variable Automatic Computer}
\acrodef{EEG}[EEG]{Electroencephalography}
\acrodef{ECG}[ECG]{Electrocardiography}
\acrodef{EMG}[EMG]{Electromyography}
\acrodef{EIN}[EIN]{Excitatory-Inhibitory Network}
\acrodef{EM}[EM]{Expectation Maximization}
\acrodef{EPSC}[EPSC]{Excitatory Post-Synaptic Current}
\acrodef{EPSP}[EPSP]{Excitatory Post-Synaptic Potential}
\acrodef{ESN}[ESN]{Echo state Network }
\acrodef{EZ}[EZ]{Epileptogenic Zone}
\acrodef{FDSOI}[FDSOI]{Fully-Depleted Silicon on Insulator}
\acrodef{FET}[FET]{Field-Effect Transistor}
\acrodef{FFT}[FFT]{Fast Fourier Transform}
\acrodef{FI}[F-I]{Frequency-Current}
\acrodef{FPGA}[FPGA]{Field Programmable Gate Array}
\acrodef{FR}[FR]{Fast Ripple}
\acrodef{FSA}[FSA]{Finite State Automaton}
\acrodef{FSM}[FSM]{Finite State Machine}
\acrodef{GIDL}[GIDL]{gate-induced-drain-leakage}
\acrodef{GOPS}[GOPS]{Giga-Operations per Second}
\acrodef{GPU}[GPU]{Graphical Processing Unit}
\acrodef{GUI}[GUI]{Graphical User Interface}
\acrodef{HAL}[HAL]{Hardware Abstraction Layer}
\acrodef{HFO}[HFO]{High Frequency Oscillation}
\acrodef{HH}[H\&H]{Hodgkin \& Huxley}
\acrodef{HMM}[HMM]{Hidden Markov Model}
\acrodef{HCS}[HCS]{High-Conductive State}
\acrodef{HRS}[HRS]{High-Resistive State}
\acrodef{HR}[HR]{Human Readable}
\acrodef{HSE}[HSE]{Handshaking Expansion}
\acrodef{HW}[HW]{Hardware}
\acrodef{ICT}[ICT]{Information and Communication Technology}
\acrodef{IC}[IC]{Integrated Circuit}
\acrodef{IEEG}[iEEG]{intracranial electroencephalography}
\acrodef{IF2DWTA}[IF2DWTA]{Integrate \& Fire 2--Dimensional WTA}
\acrodef{IFSLWTA}[IFSLWTA]{Integrate \& Fire Stop Learning WTA}
\acrodef{IF}[I\&F]{Integrate-and-Fire}
\acrodef{IMU}[IMU]{Inertial Measurement Unit}
\acrodef{INCF}[INCF]{International Neuroinformatics Coordinating Facility}
\acrodef{INI}[INI]{Institute of Neuroinformatics}
\acrodef{IO}[I/O]{Input/Output}
\acrodef{IPSC}[IPSC]{Inhibitory Post-Synaptic Current}
\acrodef{IPSP}[IPSP]{Inhibitory Post-Synaptic Potential}
\acrodef{IP}[IP]{Intellectual Property}
\acrodef{ISI}[ISI]{Inter-Spike Interval}
\acrodef{IoT}[IoT]{Internet of Things}
\acrodef{JFLAP}[JFLAP]{Java - Formal Languages and Automata Package}
\acrodef{LEDR}[LEDR]{Level-Encoded Dual-Rail}
\acrodef{LFP}[LFP]{Local Field Potential}
\acrodef{LIF}[LIF]{Leaky Integrate and Fire}
\acrodef{LLC}[LLC]{Low Leakage Cell}
\acrodef{LNA}[LNA]{Low-Noise Amplifier}
\acrodef{LPF}[LPF]{Low Pass Filter}
\acrodef{LCS}[LCS]{Low-Conductive State}
\acrodef{LRS}[LRS]{Low-Resistive State}
\acrodef{LSM}[LSM]{Liquid State Machine}
\acrodef{LTD}[LTD]{Long Term Depression}
\acrodef{LTI}[LTI]{Linear Time-Invariant}
\acrodef{LTP}[LTP]{Long Term Potentiation}
\acrodef{LTU}[LTU]{Linear Threshold Unit}
\acrodef{LUT}[LUT]{Look-Up Table}
\acrodef{LVDS}[LVDS]{Low Voltage Differential Signaling}
\acrodef{MCMC}[MCMC]{Markov-Chain Monte Carlo}
\acrodef{MEMS}[MEMS]{Micro Electro Mechanical System}
\acrodef{MFR}[MFR]{Mean Firing Rate}
\acrodef{MIM}[MIM]{Metal Insulator Metal}
\acrodef{MLP}[MLP]{Multilayer Perceptron}
\acrodef{MOSCAP}[MOSCAP]{Metal Oxide Semiconductor Capacitor}
\acrodef{MOSFET}[MOSFET]{Metal Oxide Semiconductor Field-Effect Transistor}
\acrodef{MOS}[MOS]{Metal Oxide Semiconductor}
\acrodef{MRI}[MRI]{Magnetic Resonance Imaging}
\acrodef{NDFSM}[NDFSM]{Non-deterministic Finite State Machine} 
\acrodef{ND}[ND]{Noise-Driven}
\acrodef{NEF}[NEF]{Neural Engineering Framework}
\acrodef{NHML}[NHML]{Neuromorphic Hardware Mark-up Language}
\acrodef{NIL}[NIL]{Nano-Imprint Lithography}
\acrodef{NMDA}[NMDA]{N-Methyl-D-Aspartate}
\acrodef{NME}[NE]{Neuromorphic Engineering}
\acrodef{NN}[NN]{Neural Network}
\acrodef{NRZ}[NRZ]{Non-Return-to-Zero}
\acrodef{NSM}[NSM]{Neural State Machine}
\acrodef{OR}[OR]{Operating Room}
\acrodef{OTA}[OTA]{Operational Transconductance Amplifier}
\acrodef{PCB}[PCB]{Printed Circuit Board}
\acrodef{PCHB}[PCHB]{Pre-Charge Half-Buffer}
\acrodef{PCM}[PCM]{Phase Change Memory}
\acrodef{PE}[PE]{Processing Element}
\acrodef{PFA}[PFA]{Probabilistic Finite Automaton}
\acrodef{PFC}[PFC]{prefrontal cortex}
\acrodef{PFM}[PFM]{Pulse Frequency Modulation}
\acrodef{PR}[PR]{Production Rule}
\acrodef{PSC}[PSC]{Post-Synaptic Current}
\acrodef{PSP}[PSP]{Post-Synaptic Potential}
\acrodef{PSTH}[PSTH]{Peri-Stimulus Time Histogram}
\acrodef{QDI}[QDI]{Quasi Delay Insensitive}
\acrodef{RAM}[RAM]{Random Access Memory}
\acrodef{RDF}[RDF]{random dopant fluctuation}
\acrodef{RELU}[ReLu]{Rectified Linear Unit}
\acrodef{RLS}[RLS]{Recursive Least-Squares}
\acrodef{RMSE}[RMSE]{Root Mean Squared-Error}
\acrodef{RMS}[RMS]{Root Mean Squared}
\acrodef{RNN}[RNN]{Recurrent Neural Networks}
\acrodef{RSNN}[RSNN]{Recurrent Spiking Neural Network}
\acrodef{ROLLS}[ROLLS]{Reconfigurable On-Line Learning Spiking}
\acrodef{RRAM}[RRAM]{Resistive Random Access Memory}
\acrodef{R}[R]{Ripples}
\acrodef{SAC}[SAC]{Selective Attention Chip}
\acrodef{SAT}[SAT]{Boolean Satisfiability Problem}
\acrodef{SCX}[SCX]{Silicon CorteX}
\acrodef{SD}[SD]{Signal-Driven}
\acrodef{SEM}[SEM]{Spike-based Expectation Maximization}
\acrodef{SLAM}[SLAM]{Simultaneous Localization and Mapping}
\acrodef{SNN}[SNN]{Spiking Neural Network}
\acrodef{SNR}[SNR]{Signal to Noise Ratio}
\acrodef{SOC}[SOC]{System-On-Chip}
\acrodef{SOI}[SOI]{Silicon on Insulator}
\acrodef{SP}[SP]{Separation Property}
\acrodef{SHD}[SHD]{Spiking Heidelberg Digit}
\acrodef{SRAM}[SRAM]{Static Random Access Memory}
\acrodef{SRNN}[SRNN]{Spiking Recurrent Neural Network}
\acrodef{STDP}[STDP]{Spike-Timing Dependent Plasticity}
\acrodef{STD}[STD]{Short-Term Depression}
\acrodef{STP}[STP]{Short-Term Plasticity}
\acrodef{STT-MRAM}[STT-MRAM]{Spin-Transfer Torque Magnetic Random Access Memory}
\acrodef{STT}[STT]{Spin-Transfer Torque}
\acrodef{SW}[SW]{Software}
\acrodef{TCAM}[TCAM]{Ternary Content-Addressable Memory}
\acrodef{TFT}[TFT]{Thin Film Transistor}
\acrodef{TPU}[TPU]{Tensor Processing Unit}
\acrodef{USB}[USB]{Universal Serial Bus}
\acrodef{VHDL}[VHDL]{VHSIC Hardware Description Language}
\acrodef{VLSI}[VLSI]{Very Large Scale Integration}
\acrodef{VOR}[VOR]{Vestibulo-Ocular Reflex}
\acrodef{WCST}[WCST]{Wisconsin Card Sorting Test}
\acrodef{WTA}[WTA]{Winner-Take-All}
\acrodef{XML}[XML]{eXtensible Mark-up Language}
\acrodef{CTXCTL}[CTXCTL]{CortexControl}
\acrodef{divmod3}[DIVMOD3]{divisibility of a number by three}
\acrodef{hWTA}[hWTA]{hard Winner-Take-All}
\acrodef{sWTA}[sWTA]{soft Winner-Take-All}
\acrodef{APMOM}[APMOM]{Alternate Polarity Metal On Metal}
\acrodef{SRNN}[SRNN]{Spiking Recurrent Neural Networks}
\acrodef{fMRI}[fMRI]{functional Magnetic Resonance Imaging}
\acrodef{RL}[RL]{Reinforcement Learning}
\acrodef{ES}[ES]{Evolutionary Strategies}
\acrodef{SSM}[SSM]{State-Space Model}
\acrodef{RAF}[RAF]{Resonate-and-Fire}
\acrodef{lRAF}[lRAF]{linear-Resonate-and-Fire}
\acrodef{PVT}[PVT]{Process–Voltage–Temperature}
\acrodef{TCA}[TCA]{Transconductance Amplifier}
\acrodef{SC}[SC]{Switched-Capacitor}
\acrodef{MC}[MC]{Monte Carlo}

\title{A Linear Implementation of an Analog Resonate-and-Fire Neuron\\
\thanks{The presented work has received funding from the Swiss National Science Foundation Starting Grant Project UNITE (TMSGI2-211461).}
}

\author{\IEEEauthorblockN{%
Angqi Liu\textsuperscript{*,1},
Filippo Moro\textsuperscript{*,1},
Sebastian Billaudelle\textsuperscript{1},
Melika Payvand\textsuperscript{1}
}
\IEEEauthorblockA{%
\textit{\textsuperscript{1}Institute of Neuroinformatics, University of Zurich and ETH Zurich, Zurich, Switzerland}
}
\IEEEauthorblockA{%
\textit{\textsuperscript{*}Equal contribution}
}
\IEEEauthorblockA{%
\textsuperscript{}Email: {$\{$filippo, angqi, sebastian, melika$\}$@ini.uzh.ch
}}
}

\maketitle

\begin{abstract}
Oscillatory dynamics have recently proven highly effective in machine learning (ML), particularly through \acp{SSM} that leverage structured linear recurrences for long-range temporal processing. \ac{RAF} neurons capture such oscillatory behavior in a spiking framework, offering strong expressivity with sparse event-based communication. While early analog RAF circuits employed nonlinear coupling and suffered from process sensitivity, modern ML practice favors linear recurrence. In this work, we introduce a resonate-and-fire (RAF) neuron, built in 22~nm FDSOI technology, that aligns with SSM principles while retaining the efficiency of spike-based signaling. We analyze its dynamics, linearity, and resilience to \ac{PVT} variations, and evaluate its power, performance, and area (PPA) trade-offs. We map the characteristics of our circuit into a system-level simulation where our RAF neuron is utilized in a keyword-spotting task, showing that its non-idealities do not hinder performance. Our results establish RAF neurons as robust, energy-efficient computational primitives for neuromorphic hardware.
\end{abstract}


\section{Introduction}

Oscillatory dynamics have recently emerged as a powerful computational principle in machine learning \cite{Rusch2024-mt, Effenberger2025-gg, Baronig2024-qi, Fabre2025-ta}. 
A central development in this direction has been \ac{SSM} architectures \cite{Gu2021-hn, Meyer2024-us}, which employ structured linear recurrences to capture long-range temporal dependencies with remarkable efficiency and scalability. 
By leveraging oscillatory latent dynamics, these models have demonstrated state-of-the-art performance on sequential tasks while maintaining robustness and favorable memory–computation trade-offs \cite{Rusch2024-mt}. This progress highlights the role of oscillations not only in biological systems \cite{Dogonasheva2025-gm}, but also in artificial architectures, where they act as compact and expressive primitives for temporal processing.

Introduced as a second-order extension of leaky-integrate-and-fire models, with only one state variable, the membrane state, \ac{RAF} neurons naturally embody oscillatory dynamics, thanks to the coupling of two state variables in the spiking domain \cite{izhikevich2001resonate}. Such neurons exhibit damped subthreshold oscillations, coincidence detection, and frequency selectivity. These properties render them highly expressive units for neuromorphic computing, with temporal richness and event-driven communication. 
As such, they represent a biologically inspired, yet hardware-efficient bridge between the continuous oscillations exploited in \acp{SSM} and the discrete signaling of spiking networks.

Early analog designs for \ac{RAF} neurons relied on nonlinearly coupling the two state variables, often inspired by Lotka–Volterra oscillator circuits \cite{nakada2006analog, nakada2005silicon}, to approximate the second-order dynamics of the \ac{RAF} model. While effective in demonstrating resonant behavior, these analog realizations suffered from sensitivity to device mismatch, process variations, and temperature fluctuations. Importantly, they cannot directly model the \ac{SSM}-like equations, due to the non-linear coupling of the states.
More recently, digital implementations have been explored \cite{le2023digitalRAF}, motivated by their scalability, programmability, and resilience to \ac{PVT} variations.

In this work, we revisit the \ac{RAF} neuron through the lens of modern ML-driven, i.e. \acp{SSM}, design principles. Specifically, we introduce an analog electronics implementation of a \ac{RAF}, based on 22\,nm \ac{FDSOI} technology, shown in Fig.~\ref{fig:liRAF}. 
We use \acp{TCA} to linearly couple the two states of the neuron, and leverage \ac{SC} circuits for controlling the dynamics of each state (Fig.~\ref{fig:liRAF}). Our design is energy efficient and the proposed \ac{RAF} neuron consumes [$1.6$ - $132.6$]~nW depending on the desired resonance frequency, which is set by the \ac{TCA} biasing current.
We evaluate the linearity of the coupling by analyzing the \ac{TCA}, as well as the range of the temporal dynamics provided by the \ac{SC} circuit, and analyze the non-idealities introduced by our design. 
We propose to mitigate such non-idealities with a co-design approach where the circuit's characteristics are accounted for in system-level simulation. We demonstrate the effectiveness of such an approach to solve a keyword spotting task in simulation.

\section{From SSMs to RAF neurons}

\acp{SSM} have recently demonstrated remarkable capabilities in processing long temporal sequences \cite{Gu2021-hn, Rusch2024-mt}, owing to their structured linear recurrence that captures rich dynamics while remaining computationally efficient. The continuous-time SSM formulation can be written as $\dot{x}(t)=Ax(t) + Bu(t)$ and $z(t)=\sigma( Cx(t) )$. $A$, $B$, and $C$ are the system matrices governing the state transition and readout, $\sigma$ a non-linear function.

The \ac{RAF} neuron \cite{izhikevich2001resonate} can be viewed as a biologically inspired, energy-efficient distillation of the \ac{SSM} model, embedding oscillatory state dynamics within a spiking framework. The \ac{RAF} neuron simplifies the \ac{SSM} equation by removing the $C$ matrix projection and using a heaviside non-linearity for the output. Furthermore, the \ac{RAF} model features a particular structure for the $A$ transition matrix, leading to the following continuous-time form:

\begin{equation}
    \begin{aligned}
        \dot{x}(t) &= A_{RAF}~x(t) + B_{RAF}~I(t) \\
        &= \begin{bmatrix}
            -\,\frac{1}{ \color{blue}{\tau_u} } & -\, \color{ForestGreen}{\omega_v} \\[4pt]
            \color{violet}{\omega_u} & -\,\frac{1}{ \color{blue}{\tau_v} }
        \end{bmatrix}
        \begin{bmatrix}
            u(t) \\[4pt] v(t)
        \end{bmatrix} +
        \begin{bmatrix}
            1 \\[4pt] 0
        \end{bmatrix}
        \begin{bmatrix}
            I_u(t) \\[4pt] 0
        \end{bmatrix}\\
        z(t) &= \mathcal{H}(v(t)- \Theta)
    \end{aligned}
    \label{eq_RAF}
\end{equation}
where $u$ and $v$ denote the two internal state variables (analogous to the real and imaginary parts of an SSM state $x$), $\tau_{u/v}$ is a time constant for either the $u$ or $v$ state, $\omega_{u/v}$ is the coupling factor between the neurons states, and proportional to the resonance frequency, $I(t)$ the input current. The neuron emits a spike when the state $v$ reaches a threshold $\Theta$, and we do not include a reset mechanism in the model. The circuit implementation of the Heaviside function is realized with a voltage comparator applied to the state $v$.\\

\section{Circuit implementation}

\begin{figure}[!t]
    \centering
    \includegraphics[width=0.99\linewidth]{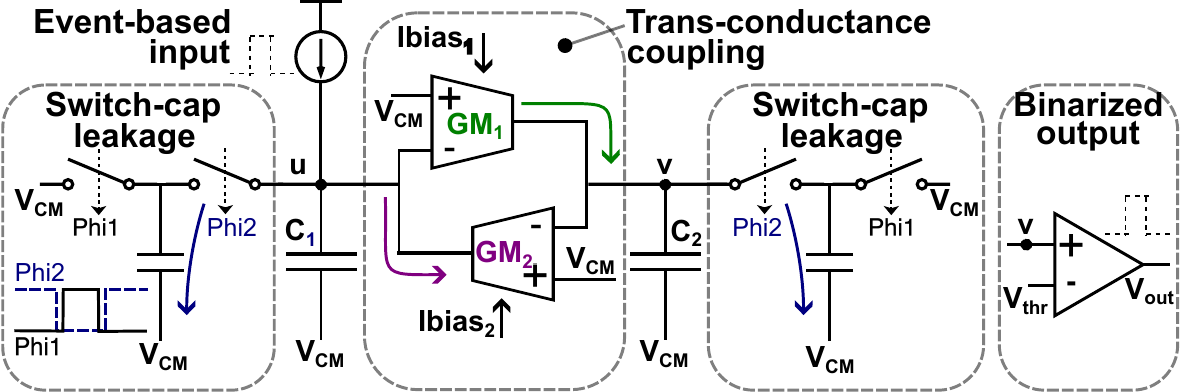}
    \caption{Linear-Resonate-and-Fire block-level circuit diagram.}
    \label{fig:liRAF}
\end{figure}

The proposed circuit realizes the \ac{RAF} neuron described in Eq.~(1), with emphasis on preserving the linearity of the state coupling coefficients. The two internal states, $u$ and $v$, are represented by voltages stored on capacitors $C_1$ and $C_2$, respectively. Their interaction follows a cross-coupling scheme mediated by a pair of linear \ac{TCA}, while their intrinsic decay is implemented through programmable \ac{SC} leakage paths. We propose the \ac{SC} instead of a DC biased subthreshold transistor as in \cite{rubino2020ultra,indiveri2011neuromorphic} to reduce mismatch and maintain fidelity to the ideal exponential decay. Figure~\ref{fig:liRAF} illustrates the overall architecture of the \ac{RAF} cell, highlighting its modular structure where oscillatory coupling and leakage are independently tunable. With reference to Eq.~\ref{eq_RAF}, the time constant is expressed as a function of the switching frequency $f_{SC}$: $\color{blue}{\tau_{u,v}}$$ = f( f_{SC} )$. Note that the circuit allows for independent control of the time constant of states $u$ and $v$ via dedicated switching frequencies~$f_{SC,u,v}$.
The resonance frequency is instead proportional to the \ac{TCA}'s linear transconductance, $g_{m1,2}$, which in turn is a function of its biasing current, $I_{bias}$: $\color{ForestGreen}{\omega_{u,v}}$$= g(I_{bias}) \approx g_{m1,2}$.

\subsection{Linear Transconductance Amplifier}
The transconductance amplifiers implement the linear cross-coupling (coefficients $\omega_{u,v}$) between the two state variables in Eq.~(1). Each amplifier converts the voltage of one state into a current that charges or discharges the other state capacitor, thereby setting the resonance frequency $\omega$. To ensure faithful reproduction of the linear \ac{SSM} dynamics, the transconductance element must exhibit high linearity over the full oscillation voltage swing, typically from ground to $V_{\mathrm{DD}}$.

\begin{figure}[!t]
    \centering\includegraphics[width=0.95\linewidth]{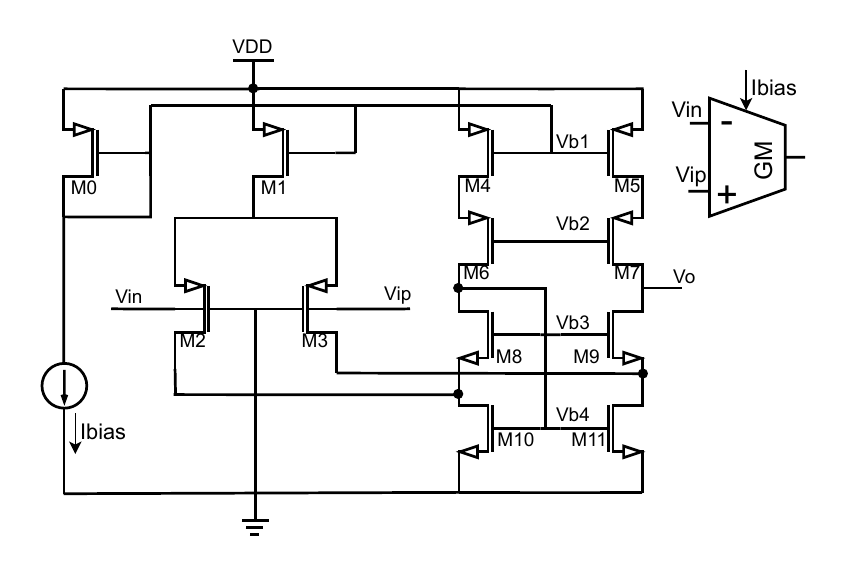}
    \caption{GM stage implemented through a bulk-driven cascoded \ac{TCA}}
    \label{fig: TCA circuit}
\end{figure}

Previous solutions for low-power, linear \acp{TCA}, such as in \cite{Sining2021-TailR}, employ a large ($\text{M}\Omega$-range) tail resistance, which occupies a large area.
In this work, this limitation is addressed by utilizing a cascoded bulk-driven differential pair \cite{BulkOTA-billy}, shown in Fig.~\ref{fig: TCA circuit}. 
This topology is selected for its ability to provide low transconductance ($g_m$), high output impedance, and high linearity. Furthermore, the implementation in \ac{FDSOI} technology is advantageous, as the bulk isolation can be more effectively realized.

As shown in Fig. 2, the amplifier's transconductance ($g_m$) is set by the input pair M2 and M3. This pair is biased by the tail current source M1, which mirrors the external reference current ($I_{bias}$) from M0. A cascode output stage (M4-M11) boosts the output impedance. Note that output impedance should be maximized to avoid non-ideal leakage from the states $u$ and $v$ through the \ac{TCA} output stages. To ensure sufficient voltage headroom for oscillation, the output stage is biased using a standard wide-swing configuration \cite{bias_design}.

\begin{figure}[t]
    \centering
    \includegraphics[width=0.9\linewidth]{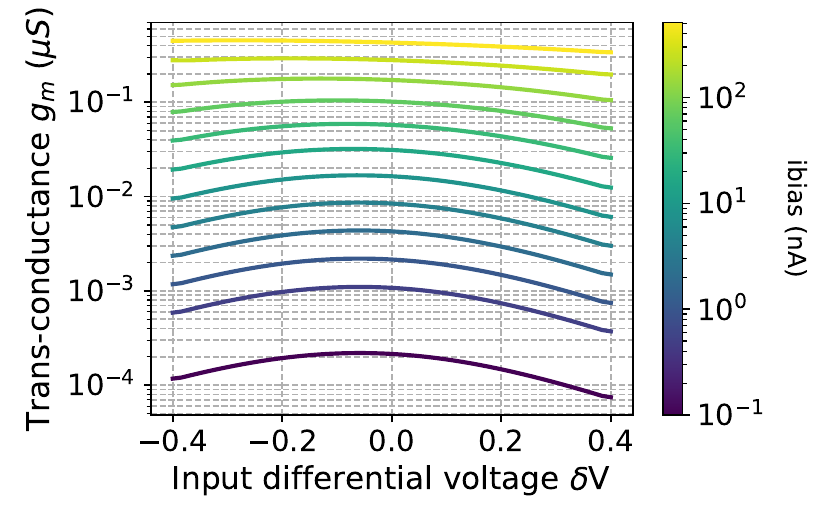}
    \caption{Linearity of the bulk-driven \ac{TCA} implementing the GM stage (Fig.\ref{fig: TCA circuit}), under different $I_{bias}$}
    \label{fig:TCA_linearity}
\end{figure}

The linearity of the proposed \ac{TCA} circuit is evaluated under various biasing conditions, with results presented in Fig.~\ref{fig:TCA_linearity}. The design achieves a relatively linear output range spanning 0 V to 0.8 V, corresponding to the full dynamic range available for the states $u$ and $v$ as constrained by the power supply rails. The variance of the transconductance ($g_m$) is subsequently employed in system-level simulations to model the non-ideal linearity of the \ac{TCA}.

\begin{figure}[t]
    \centering
    \includegraphics[width=0.99\linewidth]{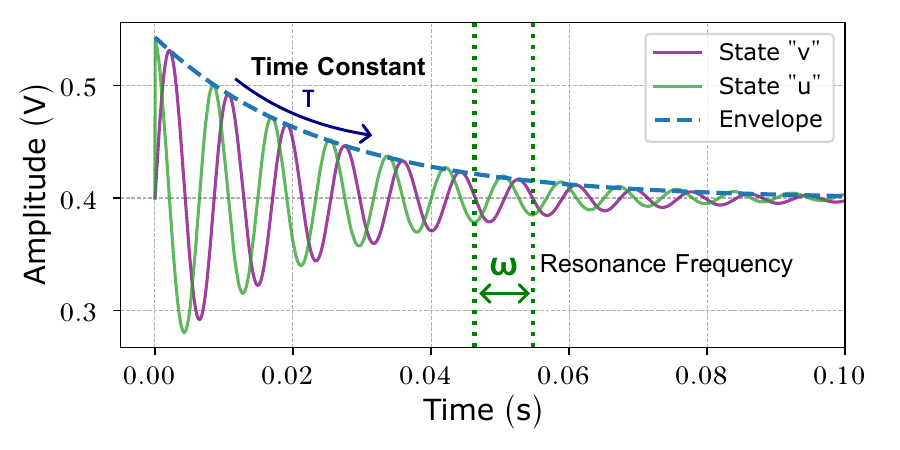}
    \caption{Dynamics of the coupled oscillator of Fig.~\ref{fig:liRAF} with $I_\text{bias}$~=~1\,pA and $f_\text{SC}$~=~10\,kHz.}
    \label{fig:oscillation dynamics}
\end{figure}

Leveraging to the linearity of the \acp{TCA}, our proposed \ac{RAF} neuron faithfully reproduces the behavior modeled by equation~\ref{eq_RAF}, as revealed by the circuit dynamics of the couple oscillator in Fig.~\ref{fig:oscillation dynamics}. This behavior is obtained using $I_\text{bias}$~=~1\,pA and $f_\text{SC}$~=~10\,kHz. 
The oscillation envelope reveals a decay behavior, governed by the \ac{SC} frequency ($f_{SC}$), in this case set to the same value for both states $u$ and $v$. Furthermore, the resonance frequency $\omega$ is determined by the \(g_m\) provided by the \ac{TCA}, which is tuned by the bias current $I_{bias}$. The coupled states $u$ and $v$ experience a \(90^{\circ}\) phase shift, and the decay aligns linearly with both the \(u\) and \(v\) signals.

\begin{figure}[!h]
    \centering
    \includegraphics[width=0.7\linewidth]{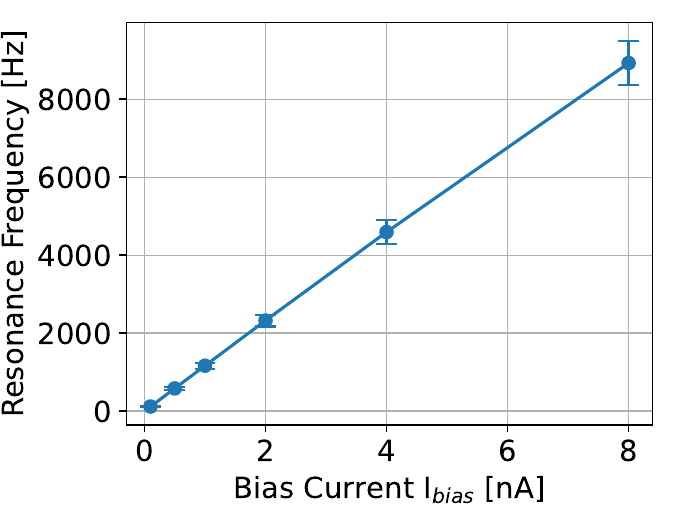}
    \caption{Resonance frequency as a function of the \ac{TCA}'s bias current.}
    \label{fig:resonance_ibias}
\end{figure}

As shown in Fig.~\ref{fig:resonance_ibias}, the resonance frequency scales with $I_{\mathrm{bias}}$ with a linear dependency, confirming that the oscillatory behavior can be adjusted through the current bias of the \ac{TCA}. This design achieves both a large dynamic range of resonance frequencies and large output impedance, the latter being critical to minimize state-dependent leakage effects discussed later.

\subsection{Switched Capacitor exponential decay}

The intrinsic decay terms of the \ac{RAF} neuron, corresponding to the diagonal elements $\tau_{u,v}$ of the system matrix in Eq.\ref{eq_RAF}, are realized using \ac{SC} circuits. Leveraging the low-leakage properties of FDSOI technology, the SC circuit can operate at a very low $f_{\mathrm{SC}}$ to achieve time constants on the order of hundreds of milli-seconds while consuming negligible power and occupying a small area. The \ac{SC} structure periodically discharges the state capacitor to a virtual common-mode node, emulating a controllable resistive path with an equivalent conductance proportional to $C_{\mathrm{fringe}} \cdot f_{\mathrm{SC}} / C_{1,2}$, where $f_{\mathrm{SC}}$ is the switching frequency. Adjusting $f_{\mathrm{SC}}$ thus allows direct tuning of the neuronal time constants $\tau_{u,v}$.

\begin{figure}[t]
    \centering
    \includegraphics[width=0.95\linewidth]{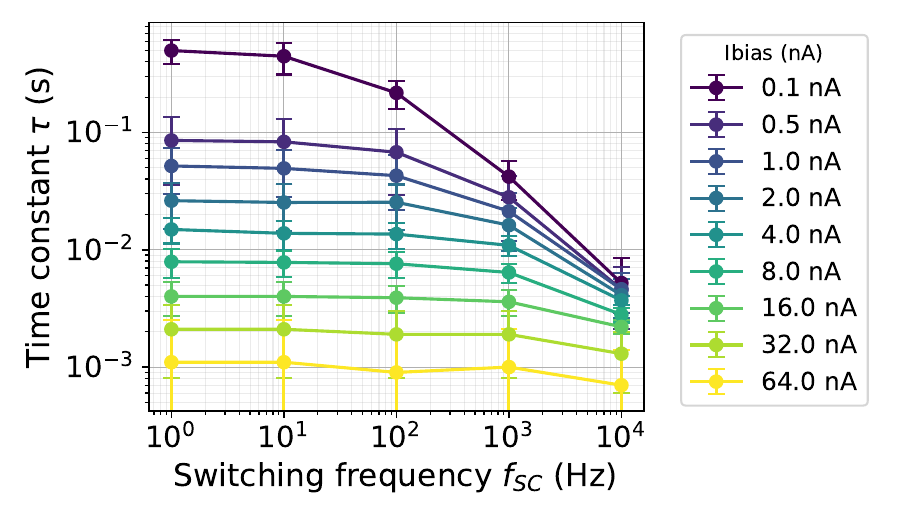}
    \caption{Time constant for the two states $u$ and $v$ as a function of the switching frequency $f_{SC}$, and for different \ac{TCA} bias currents $I_{bias}$.}
    \label{fig:time_constant_SC}
\end{figure}

\ac{MC} characterization of the circuit yields the programmable time constants reported in Fig.~\ref{fig:time_constant_SC}. 
The results show that the decay constants can be tuned over more than two orders of magnitude, approximately from $500\,\mathrm{ms}$ down to $1\,\mathrm{ms}$, covering a dynamic range compatible with temporal processing in speech, biomedical, and robotic control applications. The analysis also reveals a secondary dependency of the time constant on the bias current $I_{\mathrm{bias}}$ of the transconductance stage that drives the same node. 
This coupling between $I_{\mathrm{bias}}$ and $\tau_{u,v}$ originates from finite output impedance of the \ac{TCA} and is analyzed in the following subsection.

\subsection{Limitations of this approach}

Although the proposed design enforces linear state coupling, the finite output impedance of the transconductance amplifiers introduces an additional parasitic leakage path that modulates the effective time constant of each state. Consequently, the decay rates are not solely determined by the \ac{SC} circuitry but also by the \ac{TCA}'s output impedance, set by its DC operating point. Higher $I_{\mathrm{bias}}$ values increase the transconductance and resonance frequency while reducing output resistance, leading to a saturation, or ``clipping'', of the maximum achievable $\tau_{u,v}$, as shown in Fig.~\ref{fig:time_constant_SC}.

We mitigate this effect through two complementary strategies. First, as already mentioned, the amplifier output stages adopt a cascoded topology to boost output impedance without sacrificing transconductance linearity. Second, at the system level, we incorporate this non-ideality in a compact behavioral model of the \ac{RAF} circuit, which is used in large-scale neuromorphic simulations to accurately capture the bias-dependent time-constant compression. This co-design approach ensures that algorithmic simulations and circuit-level performance remain consistent.

\subsection{Summary of circuit performance}

Table \ref{tab:PPA osc} summarizes the performance of the designed \ac{RAF}. In contrast to previously proposed neurons \cite{nakada2005silicon}, this design provides enhanced control over a broad range of time constants and intrinsic oscillation frequencies, with a tuning range sufficient for real-time signal processing applications.

\begin{figure}[!h]
    \centering
    \includegraphics[width=0.7\linewidth]{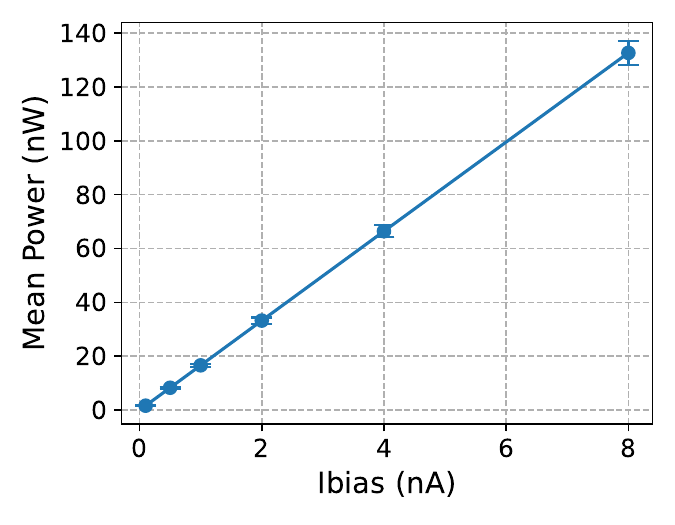}
    \caption{Dynamic power consumption as a function of the bias current $I_{bias}$}
    \label{fig:power}
\end{figure}

The total power consumption of the \ac{RAF} neuron is dominated by the static biasing of the transconductance amplifiers. Because the \ac{TCA} is current-biased, its contribution scales approximately as $I_{\mathrm{bias}} \cdot V_{\mathrm{DD}}$, leading to an almost linear relationship between oscillation frequency and power. The \ac{SC} section contributes only a small dynamic component proportional to $f_{\mathrm{SC}}$, which remains nearly constant across our power estimation. As a result (Fig.~\ref{fig:power}), the total dynamic power grows proportionally with $I_{\mathrm{bias}}$, with an empirical slope of $16.5\,\mathrm{nW}/\mathrm{nA}$. The reported range of $1.6$--$132.6\,\mathrm{nW}$ corresponds to $I_{\mathrm{bias}} \in 0.1$--$8.0\,\mathrm{nA}$. The output comparator is not included in this estimate.

\begin{table}[htbp]
\centering
\caption{Performance summary the \ac{RAF} circuit. *Depending on TCA bias current and required resonance frequency} 
\label{tab:PPA osc} 
\resizebox{0.75\columnwidth}{!}{%
\begin{tabular}{|l|c|c|c|c}
\hline
Technology [nm] & 22 \\ \hline
Oscillation principle & Gm-C \\ \hline
Oscillator type & Harmonic \\ \hline
Supply voltage [V]  & 0.8 \\ \hline
Power consumption [nW]  & $1.6 - 132.6$* \\ \hline
Frequency range [Hz] & $100 - 500 \times 10^3$ \\ \hline
Area [$\mu$m$^2$] & 6524 \\ \hline
Additional voltage regulator/ & NO/YES \\ 
current reference required & \\ \hline
\end{tabular}%
}
\end{table}

\section{System Level simulation of the lRAF circuit}

To evaluate the computational capabilities of the proposed \ac{RAF} circuit under realistic hardware constraints, we performed a system-level simulation on a keyword spotting task using the Spiking Heidelberg Digits (SHD) dataset. The experiments build upon the codebase in \cite{Baronig2024-qi}, adapted to match our \ac{RAF} dynamics and parameterization. All simulations employ 8-bit quantized synaptic weights, reflecting the precision typically used in edge neuromorphic processors where memory footprint is a critical constraint.

\begin{figure}[t]
    \centering
    \includegraphics[width=0.72\linewidth]{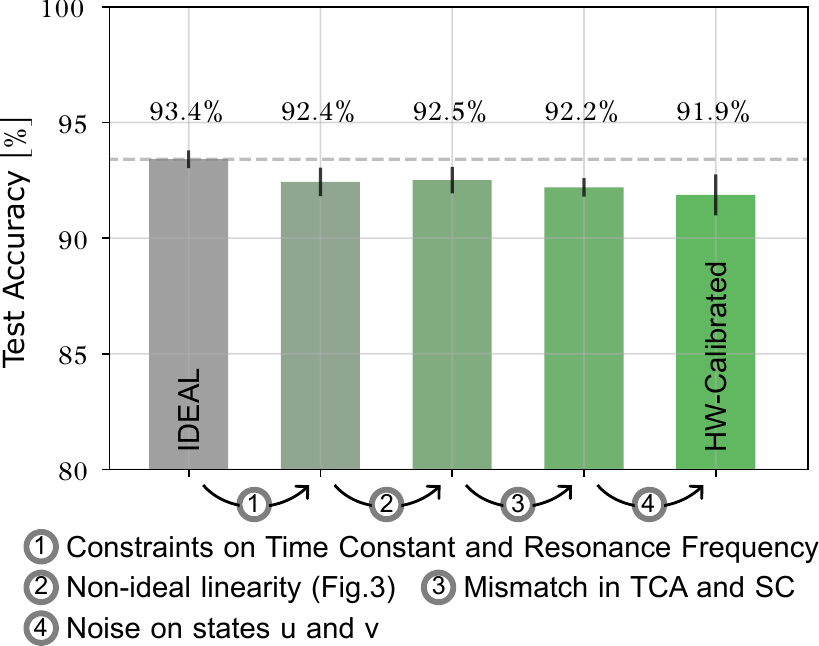}
    \caption{Classification accuracy on the SHD task with different levels of hardware-awareness to the \ac{RAF} circuit's non idealities.}
    \label{fig:SHD}
\end{figure}

We progressively introduce different degrees of hardware awareness into the neuron model to bridge the gap between the ideal mathematical formulation in Eq.~(1) and simulation from the designed circuit. We estimated fixed-pattern deviations and temporal noise from Monte Carlo and transient noise simulations. Five levels of abstraction are defined.
First, the Ideal model, which is the discretized version of the continuous-time \ac{RAF} equations, without any hardware constraints. Progressively, we add: (1) Constraints on the resonance frequency $\omega$ and time constants $\tau_{u,v}$ according to Fig.~\ref{fig:resonance_ibias} and Fig.~\ref{fig:time_constant_SC}; (2) Non-ideal linearity of the \ac{TCA} according to Fig.~\ref{fig:TCA_linearity}; (3) Device Mismatch acting on the resonance frequency $\omega$ and time constants $\tau_{u,v}$ parameters; (4) Additive white noise on the states $u$ and $v$, esulting in a complete hardware-calibrated behavioral model.
The classification results across these levels are summarized in Fig.~\ref{fig:SHD}, showing a small drop (1.5\%) of classification accuracy from the ideal to the hardware aware level. 


\section{Conclusion}
We present a \ac{RAF} circuit that follows the recent oscillatory models in machine learning, such as \acp{SSM}, by implementing a linear two-state coupling. The circuit employs a bulk-driven transconductance amplifier and a switched-capacitor design, achieving both high linearity and time constants suitable for real-time signal processing. While intrinsic non-idealities introduce deviations from the ideal linear \ac{RAF} behavior, their impact is mitigated through hardware-aware system-level simulations. In this work, we prove the efficacy of this co-design approach achieving competitive keyword spotting performance. Future work will be needed to extend the approach to different task domains. 


\printbibliography










\end{document}